\documentstyle[12pt]{article}

\input{tcilatex}

\begin{document}

\begin{titlepage}
\begin{center}
\null
{\bf \large Vanishing Loss Effect on the Effective $ac$ Conductivity
behavior for  2D Composite Metal-Dielectric Films At The Percolation
Threshold}\\
\vspace{1.5cm}
L. Zekri and N.Zekri \footnote{ Regular Associate of the ICTP.}\\ 
{\em U.S.T.O., D\'{e}partement de Physique, L.E.P.M.,\\  B.P.1505 El 
M'Naouar, Oran, Algeria\\
and\\
The Abdus Salam International Centre for Theoretical Physics,  Trieste, Italy,}\\
\bigskip
{\em and}\\
\bigskip
J.P. Clerc\\
{\em Ecole Polytechnique Universitaire de Marseille, Technopole Chateau
Gombert, 5 Rue Enrico Fermi 13453, Marseille France. } \\
\vspace{2cm}
MIRAMARE -- TRIESTE\\
\medskip
\end{center}
\end{titlepage}

\centerline{\bf Abstract} \baselineskip=20pt

\hspace{0.33in} We study the imaginary part of the effective
$ac$ conductivity as well as its distribution probability for vanishing
losses in 2D composites. This investigation showed that the effective
medium theory provides only informations about the average conductivity,
while its fluctuations which correspond to the field
energy in this limit are neglected by this theory. \newline

\newpage

\noindent{\bf Introduction }

\hspace{0.33in} Although metal-dielectric composites were extensively
investigated during the last decade, some cases where the local field and
conductivity fluctuations are strong remain not understood \cite{Brou1,Zek1}%
. The effective medium theory \cite{Medium} relates at the percolation
threshold, the effective conductivity to the metal and dielectric
conductivities by

\begin{equation}
\sigma_{eff} =\sqrt{\sigma_{m} \sigma_{d}}  \label{effective}
\end{equation}

The indices $m$, $d$ and $eff$ stand respectively for the metal, dielectric
and effective medium. However, this equation is not satisfied for vanishing
loss, where it expects the effective $ac$ conductivity to be real while
numerical studies showed the real part to be vanishing \cite{Zek1}.
Brouers et al. \cite{Brou2}
explained the field behavior in this limiting case as an energy storage of
the electromagnetic wave between dielectric and superconducting
component. In the same sens, we have found the local field to be localized
in this limit \cite{Zek2}. This localization is enhanced by
the dissipation for finite losses.

\hspace{0.33in} However, no work (to the best of our knowldge) has been
devoted to the imaginary part of the effective conductivity in this limit of
the loss. This is the aim of the present work where we study the loss effect
on the imaginary part of the effective conductivity as well as its
distribution in its vanishing limit for a 2D metal-dielectric network.

\noindent{\bf Model description }

\hspace{0.33in} The metal-dielectric composites can be modelled as an $RLC$
network where the metallic grains represent the $RL$ branches while
dielectric grains correspond to the capacitance $C$ \cite{Zeng}. The
frequency dependent conductivities behave then respectively for the
metal and dielectric components as
\begin{equation}
\sigma_{m,d}= \frac{-i d \omega \epsilon_{m,d}}{4 \pi} \label{sigmaepsi} 
\end{equation}
\begin{equation}
\sigma_{m} = \frac{1}{i L \omega +R }
\end{equation}
\noindent and 
\begin{equation}
\sigma_{d} = i C \omega
\end{equation}

$R$ being the loss in the metal. We limit ourselves in this paper to the
characteristic frequency $\omega_{res}$ where $\left| \sigma_{m} \right| =
\left| \sigma_{d} \right|$ for vanishing loss, i.e. $L \omega_{res} = 1 / C
\omega_{res}$. The inductance $L$ and capacitance $C$ being constant near
the characteristic frequency, we can take without loss of generality
$L=C=\omega_{res}=1$. For any frequency close to $\omega_{res}$ the
frequency $\omega$ in these equations is expressed as $\omega /
\omega_{res}$. Although this form simplifies the calculations, we can
easily determine the true parameters of the dielectric constants of the two
components \cite{Zek2}.

\hspace{0.33in} In order to determine the effective conductivity of the
resulting $RLC$ network, we solve exactly the sets of Kirchoff equations
of the $RLC$ network by handling the corresponding large matrices
by blocs of small matrices due to their sparce shape and arrangement
in the diagonal region (see \cite{Zek3}).

\noindent {\bf Results and discussion}

\hspace{0.33in} In a 2D square lattice, the percolation threshold is reached
when the concentration of the metallic component is $p=0.5$ \cite{Stauff}.
For the rest of the paper, we restrict ourselves to this concentration. We
then consider an $RLC$ lattice of $256$x$256$ branches in order to ensure a
statistically stable system. We have previously found the real part of the
effective conductivity to vanish for a vanishing loss \cite{Zek2}. This
result does not agree with the predictions of the effective medium theory
(given by Eq. 1) where the imaginary part is expected to vanish while the
real one should be finite in this limit. However, in the case of very small
losses, the local field strongly fluctuates \cite{Brou1} and the
distribution of real part of the effective conductivity is Poissonian and
therefore, not characterized by its averaged value \cite{Zek1}. Such
fluctuations are not taken into account by the effective medium theory which
determines only the average effective conductivity. First, we will study the
effect of vanishing loss on the imaginary conductivity (which was not
examined in the previous works \cite{Zek1,Zek2}).

\hspace{0.33in} In figure 1, we show the loss dependance of the imaginary
part of the effective conductivity for two sample realizations. The main
feature in this figure is a transition behavior at around $R=10^{-5}$ Ohm from a
vanishing conductivity for large losses (expected from Eq.1) to finite one
for very small losses where this conductivity saturates. This transition
region (which separates the two different statistical behaviors for small
and large losses) was also observed in the behavior of the real
conductivity  \cite{Zek1}. We see also from Fig.1 that the imaginary
conductivity can have either positive or negative values depending on the
sample realization. Therefore the system should strongly fluctuate between
metallic (inductive) and dielectric (capacitive) behavior in the limit of
very small losses. This leads us to examine the statistical properties of
the imaginary effective conductivity.

\hspace{0.33in} These statistical properties are clearly shown in figure 2 
for two different sizes \ for comparison ($150x150$ and $256$x$256$) and for $%
450$ realizations with $R=10^{-9}$ Ohm, where the distribution of the imaginary
effective conductivity is a symmetric centered at $0$. The
fluctuations seem to decrease much slowly with size than the square root 
of the size (which is characteristics of normal distributions). Therefore 
this distribution may have long tails like levy distributions \cite{Levy}. These  
results agree with the prediction of Eq. 1 regarding the average imaginary 
conductivity, but in this case the main information should be provided by its 
fluctuations and not the average since the complex conductivity vanishes.
Indeed, since the real conductivity was found to vanish (its distribution
tends to a delat-peak at $0$ when the system is
very large) as
well as the average imaginary conductivity (while the components
conductance is finite). It seems that their fluctuations contain
informations on the field energy in agreement with the
previous arguments \cite{Zek1, Brou2, Zek2}. 

\noindent{\bf Conclusion }

\hspace{0.33in} In this article we have studied the imaginary part of the
effective conductivity in the limit of vanishing loss. We found it to vanish
in avergage but with strong fluctuations. These results show that the
average conductivity loses the informations about the field energy which can
be provided by the conductivity fluctuations.

\noindent{\bf Acknowledgements }

\hspace{0.33in} Two of us (L.Z and N.Z) would like to acknowledge the
hospitality of the Abdus Salam I.C.T.P. during the progress of this work. 
Financial support from the Arab Fund is Ackowledged by N.Z. We thank also Professor 
Dykhne for drawing our attention about the imaginary effective conductivity.

\newpage

\begin{center}
{\bf Figure Captions}
\end{center}

\bigskip

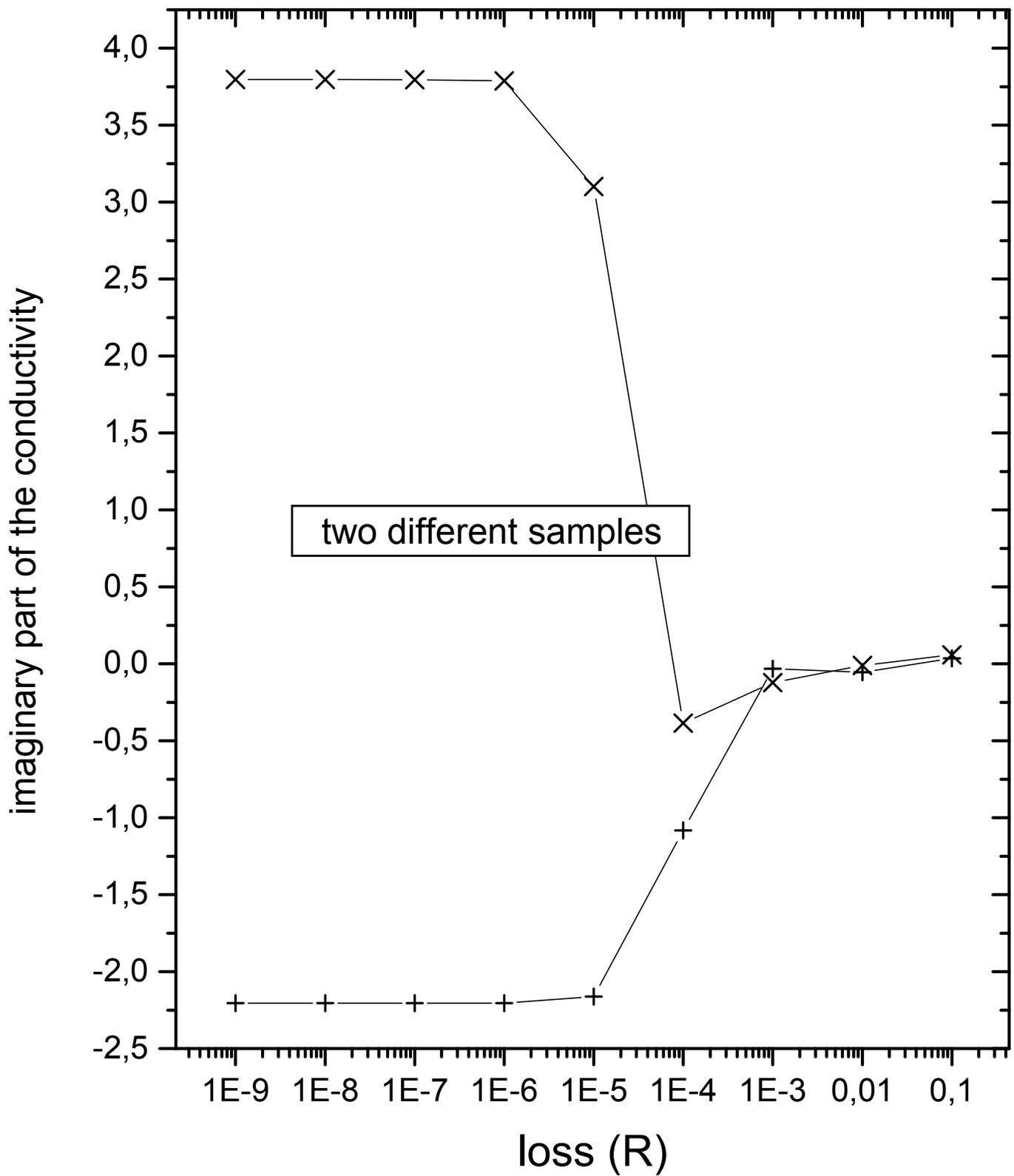
{\bf Fig.1 } The loss dependance of the imaginary part of the effective
conductance for two different realization samples.

\bigskip

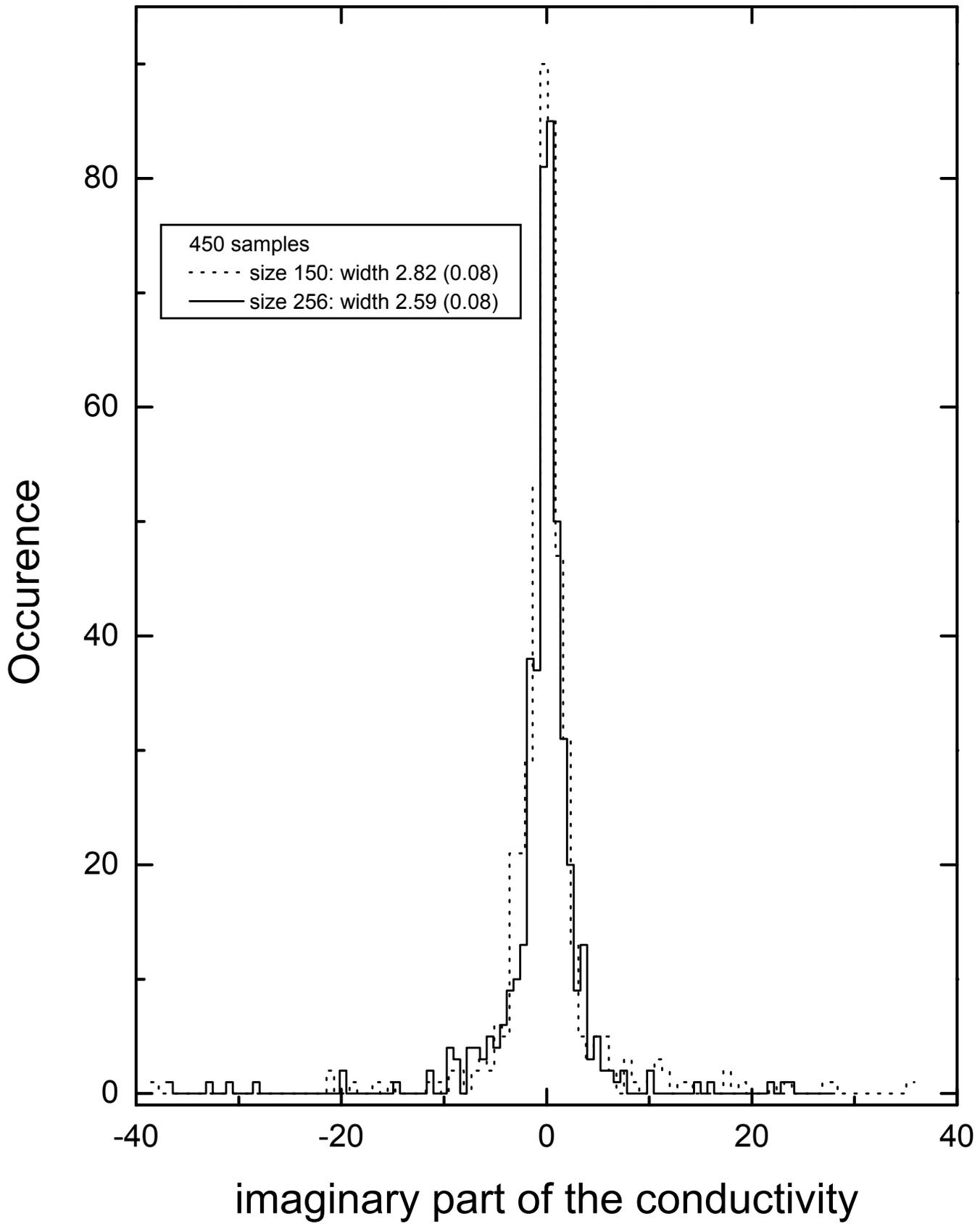
{\bf Fig.2} The distribution of the imaginary part of the effective
conductance for $450$ samples of two different sizes, the loss is
$10^{-9}$.

\bigskip

\end{document}